\documentstyle[aps,prb,multicol,eqsecnum]{revtex}

\begin{document}
\draft

\title{Effect of electron correlation \\ on phonons in a strongly-coupled
  electron-phonon system}

\author{Takashi Hotta and Yasutami Takada}

\address{Institute for Solid State Physics, University of Tokyo,
  7-22-1 Roppongi, Minato-ku, Tokyo 106, Japan}

\date{\today}

\maketitle

\begin{abstract}

An asymptotically exact result is obtained for the renormalized phonon
energy as a function of the on-site Coulomb repulsion $U_{ee}$ in the
half-filled Hubbard-Holstein model in the strong-coupling region at
zero temperature. 
The result is obtained on the basis of the effective Hamiltonian
derived in the antiadiabatic region with due attention to the 
transition from CDW- to SDW-regime with the increase of $U_{ee}$.
Somewhat contrary to naive expectation, the phonon energy is found
to shift in the insulating phase and the shift becomes larger
in the SDW regime than that in the CDW regime.

\end{abstract}

\pacs{PACS numbers: 71.38.+i, 71.10.Fd, 71.28.+d, 63.20.-e}

\begin{multicols}{2}
\narrowtext


\section{Introduction}
\label{sec:1}

For more than two decades, organic materials have attracted special
attention due to the interesting features of a complex system in 
which charge, spin, and phonon degrees of freedom correlate to one
another.\cite{organic}
Similar complex features have emerged recently in the ferromagnetic
perovskite compounds La$_{1-x}A_x$MnO$_3$ where $A$ represents a
divalent alkaline-earth element such as Ca or Sr.\cite{CMR1,CMR2,CMR3} 
Many people also anticipate that the mechanism of high-$T_c$
superconductivity in the copper oxides cannot be fully understood
without a viewpoint to regard these materials as a
charge-spin-phonon complex system.\cite{HTC1,HTC2,HTC3}

The Hubbard-Holstein (HH) model provides a useful framework to
investigate such a complex system.
Besides the kinetic-energy term represented by the nearest-neighbor
electron hopping integral $t$, it consists of two interaction parts.
One is the term to treat the electron correlation effects through
the short-range Coulomb repulsion $U_{ee}$ as in the Hubbard model.
\cite{hubbard}
The other is the term to couple an electron with local phonons,
leading to the phonon-mediated electron-electron attraction
$-U_{ph}$.
The model without the $U_{ee}$ term is called the Holstein model
which was proposed to consider the motion of a conduction electron
coupled with intramolecular vibrations of ions in a molecular
crystal. \cite{holstein}
This Holstein model has a long history of research and in the
last several years it has been a subject of intensive studies
based mainly on the quantum Monte Carlo (QMC) simulations.
\cite{scalapino,marsiglio1,freericks1,freericks2}

In the adiabatic region in which $t$ is much larger than
the local-phonon energy $\omega_0$, the HH model as well as
the Holstein model can be studied with use of the traditional
Migdal-Eliashberg (ME) theory. \cite{migdal,eliashberg}
In the antiadiabatic region ($t \lesssim \omega_0$), however,
a more careful treatment is necessary and numerical approaches
such as an exact diagonalization method
\cite{ranninger1,alexandrov1,zhong,marsiglio2,takada1}
and QMC simulations \cite{freericks3,berger}
have been employed.
These studies have revealed that the HH model can afford
interesting phase diagrams representing the competition
among charge density wave (CDW), spin density wave (SDW),
and superconductivity as we change the parameters involved
in the system, i.e., $U_{ee}$, $U_{ph}$, $t$, $\omega_0$,
and $n$ the electron filling.
Crudely speaking, the CDW (SDW) nature manifests itself
in the ground state at half filling, if the effective
electron-electron interaction $U$ defined by
$U \equiv U_{ee}-2U_{ph}$ is negative (positive).

In this paper, we focus on the renormalized local-phonon energy
$\tilde{\omega}_0$ in the HH model in the antiadiabatic region
at zero temperature ($T=0$) and inquire how much change can be
seen in $\tilde{\omega}_0$ at the CDW-SDW transition.
Note that the shift $\delta \omega_0$ defined by
$\delta \omega_0 \equiv \omega_0 - \tilde{\omega}_0$ is
directly connected with density fluctuations the amount of which
depends sensitively on the nature of the ground state.
We shall give an exact result for $\delta \omega_0$ as a function
of $U_{ee}$ in the strong-coupling region ($U_{ph} \gg \omega_0$)
and discuss its change at the CDW-SDW regime boundary.
We consider the half-filling situation at which we presume
that the largest shift in $\tilde{\omega}_0$ appears.

The antiadiabatic condition is usually fulfilled in the
strong-coupling region 
due to the small-polaronic band-narrowing effect.
This is also emphasized by 
Alexandrov and Capellmann (AC) \cite{alexandrov2,alexandrov3} who,
starting with a comment 
on the importance of antiadiabatic motion of electrons
in a strongly-coupled electron-phonon system,
have derived an analytical result for $\delta \omega_0$
presumably at $U=0$ by using the Green's-function method.
The result of AC is obtained at $T$ larger than the
critical temperature for the ordered state of CDW or SDW.
In such a temperature range, the amount of density fluctuations
differs much from that at $T=0$.
Thus we expect to obtain a result for $\delta \omega_0$
which is different from that of AC.

Subtlety exists in the present problem as for the inclusion of
an ordering nature of electrons in the Green's-function method.
In order to avoid the difficulty originating from this subtlety,
we employ an approach based on the effective Hamiltonian
$H_{\rm eff}$ derived from the HH model in the antiadiabatic
and strong-coupling region at half filling.
We observe that $H_{\rm eff}$ for $U<0$ (the CDW regime) is
nothing but an Ising-like pseudospin Hamiltonian with an exchange
interaction depending on both $|U|$ and local phonon number.
On the other hand, $H_{\rm eff}$ for $U>0$ (the SDW regime)
can be mapped onto the Heisenberg spin Hamiltonian with the same
exchange interaction.

By evaluating $\tilde{\omega}_0$ with use of these $H_{\rm eff}$'s,
we find that $\delta \omega_0$ depends on the sign of $U$.
If we compare them at the same value of $|U|$, $\delta \omega_0$
for $U>0$ is larger than that for $U<0$.
This result may be understood if we remember the fact that phonons
with energy much larger than the characteristic energy of
small polarons or bipolarons can react to both spin and charge
fluctuations in the electronic system.
In the HH model in the strong-coupling region, spin degrees
of freedom give larger fluctuations than those of charge.

In Sec.~\ref{sec:2}, we give the Hamiltonian for the HH model
and explain an overall feature of $\tilde{\omega}_0$
as we change the parameters in the system in order to specify
the quantity to which we pay attention in this paper.
In Sec.~\ref{sec:3}, we show various effective Hamiltonians,
$H_{\rm eff}$'s, depending on the nature of the ground state
in the antiadiabatic region.
In Sec.~\ref{sec:4}, the phonon energy shift $\delta \omega_0$ is
calculated at half filling based on the $H_{\rm eff}$'s.
Finally in Sec.~\ref{sec:5}, we summarize our results and
briefly discuss a remaining problem such as the dependence on $n$.
Supplementary discussions are given in Appendices A, B, and C. 
Throughout this paper, we employ units in which $\hbar=k_{\rm B}=1$.

\section{Hubbard-Holstein model}
\label{sec:2}

Consider a system of electrons interacting with one another
through the short-range Coulomb repulsion and hopping around in a
crystal composed of molecular units, each of which provides an
intramolecular high-energy optic phonon coupled to the electrons.
Assuming that both the energy of phonon $\omega_0$ and the 
electron-phonon coupling constant are independent of wave vectors, 
we can write the Hamiltonian for the $N$-site system as
\begin{eqnarray}
  \label{hh}
  H &=& -t \sum_{i,j,\sigma} N_{ij} c_{i\sigma}^{\dagger}c_{j\sigma} 
  - \mu \sum_{i}\rho_i
  + U_{ee} \sum_i n_{i\uparrow}n_{i\downarrow}  \nonumber \\ 
  &+& \sqrt{\alpha} \omega_0 \sum_{i} \rho_i
  (a_i+a_i^{\dagger}) + \omega_0 \sum_i a_i^{\dagger}a_i, 
\end{eqnarray}
where $N_{ij}$ is defined as unity for a nearest-neighbor-site pair,
$\langle i,j \rangle$, and zero otherwise, 
$c_{i\sigma}$ the operator to annihilate a spin-$\sigma$
electron at site $i$,
$n_{i\sigma} \equiv c_{i\sigma}^{\dagger}c_{i\sigma}$, 
$\rho_i=n_{i\uparrow}+n_{i\downarrow}$, 
$\mu$ the chemical potential, 
$\alpha$ the nondimensional electron-phonon coupling constant defined
through $\alpha \equiv U_{ph}/\omega_0$, 
and $a_i$ the phonon annihilation operator at site $i$. 

Information about the phonon properties is involved in the phonon
Green's function $D({\bf q},i\omega_{\nu})$, defined by 
\begin{eqnarray}
  \label{eq:phD}
  D({\bf q}, i\omega_{\nu}) 
  &=& - \int_0^{1/T} d\tau e^{i \tau \omega_{\nu}}
  \Bigl\langle T_{\tau} \phi_{\bf q}(\tau)\phi_{-{\bf q}}
  \Bigr\rangle, 
\end{eqnarray}
where $\omega_{\nu}\equiv 2\pi \nu T$ is the boson Matsubara
frequency with an integer $\nu$, $T_{\tau}$ is the time ordering
operator, $\phi_{\bf q} \equiv a_{\bf q}+a_{-{\bf q}}^{\dagger}$
with $a_{\bf q} \equiv N^{-1/2} \sum_i e^{i{\bf q} \cdot {\bf R}_i}
a_i$ represents the ion-displacement operator, and
$\langle \cdots \rangle$ denotes the thermodynamic average.
The renormalized phonon energy $\tilde{\omega}_0$ appears as
a pole of the retarded form of $D$, $D^{\rm R}({\bf q},\omega)$,
in the complex $\omega$ plane.

If we employ the Green's-function method for the evaluation of
$D({\bf q},i\omega_{\nu})$ in the antiadiabatic ($t \ll \omega_0$)
and strong-coupling ($U_{ph} \gg \omega_0$) region, we encounter a
really complicated situation because we need to make a proper
inclusion of the vertex corrections. \cite{takada2}
On the other hand, an exact diagonalization method for the study of
a small-size HH system is useful in studying an overall qualitative
feature of $\tilde{\omega}_0$ in this region.
The size effect is expected to be small due to the infrequent
hopping nature of electrons. \cite{alexandrov1}
In Fig.~\ref{fig:1}, we show a typical result for the excitation
energies in the energy region less than $3\omega_0$
as a function of $U_{ee}$ with a fixed value of $U_{ph}$
in a two-site half-filled HH model.
We obtain this result by using an exact diagonalization method
\cite{ranninger1,takada1} to calculate the pole positions in
$D^{\rm R}(q,\omega)$ at $q=\pi$
(the antibonding molecular orbital state) with the parameters
$t=0.2\omega_0$ and $U_{ph}=2\omega_0$.

There are basically two kinds of ion displacements in this region.
One is the shift of equilibrium ion position
associated with the change in local electron number.
The other is the vibration of small amplitude around each
equilibrium position.
Accordingly, $\tilde{\omega}_0$ has two main branches of modes
in Fig.~\ref{fig:1}(a) except for $U \approx 0$.
The former mode of purely electronic origin represented
by the dashed curve was discussed by Ranninger \cite{ranninger2}
as a lattice deformation wave driven by charge fluctuation
of the system.
Thus we denote the mode as ``charge''.
(For comparison, we have also shown the spin-excitation energy
by the thin dashed curve in the figure, though it does not give
a pole structure in $D^{\rm R}$.)
The latter mode is an intrinsic phonon mode always appearing at the
energy slightly less than $\omega_0$.
We plot this mode by the solid curve and label it as ``phonon''.

The charge mode has a large spectral weight of the order of
$8\alpha$ in the region of $U<0$ (the CDW regime) as shown in
Fig.~\ref{fig:1}(b) and this is the mode which evolves from the
softened mode in the adiabatic region \cite{engelsberg}
with the increase of $U_{ph}$ at a fixed value of $U_{ee}$.
However, it does not deserve further investigation, because
in the CDW regime $\tilde{\omega}_0$ is known to be vanishingly
small, i.e., of the order of the exponentially small bipolaron
hopping energy and in the SDW regime ($U>0$) its spectral weight
is negligibly small.

The intrinsic phonon mode has the spectral weight of about unity
irrespective of $U$.
In the following, we mainly concern ourselves with the amount of
the energy shift $\delta \omega_0$ of this mode.
Note that this phonon mode couples with the charge mode at
$U_{ee} \approx 2U_{ph}+\omega_0$, originating from the
coincidence of the bare phonon energy with the charge excitation
energy at this value of $U_{ee}$.
We will leave a detailed study of this special situation
in the future.

Three comments are in order concerning the results
on the small-size HH model.
(1) For the special situation of $U \approx 0$ with not so
strong $U_{ph}$ in which a metallic state appears as a consequence
of the competition between CDW and SDW, \cite{takada1}
the third mode plotted by the dotted curve in Fig.~\ref{fig:1}
has also a considerable spectral weight.
This mode represents the coupling between charge and intrinsic
phonon modes.
For this reason, we denote this mode as ``charge-phonon''
in Fig.~\ref{fig:1}.
(2) The Green's-function approach of AC
\cite{alexandrov2,alexandrov3} involves only two-site processes.
The lattice effect is included merely through the coordination
number $z$.
Thus their result should be consistent with our two-site
calculation on $\delta \omega_0$.
We find that the parameter dependence, i.e.,
$\delta \omega_0 \propto zt^2/(\alpha^2 \omega_0)$, agrees,
but the proportionality constant is different.
We also note that according to our rigorous argument
in Sec.~\ref{sec:4}, $\delta \omega_0$ depends on the lattice
structure in a more complicated way than merely through $z$
even in the strong-coupling limit.
This is especially the case for the SDW ground state.
(3) A rather extensive QMC calculation was done by Berger
{\it et al.} on the $4 \times 4$ two-dimensional (2D) HH model.
Their result on $\tilde{\omega}_0$ as a function of $U_{ee}$
is given in Fig.~8 in Ref.~\onlinecite{berger} for the
half-filled system at ${\bf q}=(\pi,\pi)$ and with the values
of $U_{ph}=t=\omega_0$ and $T=0.1\omega_0$. \cite{berger2}
It seems that they have detected only the dominant mode
in each region, i.e., the charge mode for $U<0$ and
the phonon mode for $U>0$.
With such an interpretation of their QMC result, we find that they
have obtained a very similar result to that in Fig.~\ref{fig:1}.
This is encouraging, because the theory for $\tilde{\omega}_0$
in the strong-coupling and antiadiabatic region seems to be
applicable even to the medium-coupling,
i.e., $U_{ph} \approx \omega_0$, and not very antiadiabatic,
i.e., $t \approx \omega_0$, case.

\section{Effective Hamiltonian in the antiadiabatic region}
\label{sec:3}

\subsection{Lang-Firsov transformation}
\label{sec:3A}

The Lang-Firsov (LF) canonical transformation \cite{lang} is useful
to treat the HH model in the strong-coupling region.
It converts $H$ into $\bar{H}$ as
\begin{equation}
  \label{eq:Hbar}
  \bar{H} = e^P H~e^{-P} = \bar{T} + \bar{V},
\end{equation}
where $P$ is defined as $P = -P^{\dagger} = 
\sqrt{\alpha}\sum_{i}\rho_{i}(a_i^{\dagger}-a_i)$,
$\bar{T}$ is the inter-site hopping term, described as
\begin{eqnarray}
  \label{eq:hopping}
  \bar{T} = -t \sum_{i,j,\sigma} N_{ij}
  c_{i\sigma}^{\dagger}c_{j\sigma} X_{i}^{\dagger}X_{j},
\end{eqnarray}
with $X_i={\rm exp}[\sqrt{\alpha}(a_i-a_i^{\dagger})]$,
and $\bar{V}$ the on-site term is given by
\begin{eqnarray}
  \label{eq:on-site}
  \bar{V} = U \sum_{i}n_{i\uparrow}n_{i\downarrow}
  +\omega_0\sum_{i}a_i^{\dagger} a_i
  -(\mu+U_{ph})\sum_{i}\rho_{i}.
\end{eqnarray}
Note that the on-site instantaneous interaction is changed from
$U_{ee}$ into $U=U_{ee}-2U_{ph}$.
Note also that $c_{i\sigma}$ in $\bar{H}$ should be interpreted as
an annihilation operator of a ``polaron'', i.e., an electron
dressed with ion displacements, rather than a bare electron in $H$.
Similarly, $a_i$ in $\bar{H}$ represents an annihilation operator
for the quantized ion vibration around $\sqrt{\alpha}\rho_i$
the shifted equilibrium position depending on local electron number.
In accordance with this change of $a_i$,
$D({\bf q},i\omega_{\nu})$ in Eq.~(\ref{eq:phD}) can be decomposed
into three parts as
\begin{eqnarray}
  \label{eq:phD2}
  D({\bf q}, i\omega_{\nu}) &=& D_{ph}({\bf q},i\omega_{\nu})
  - 2\sqrt{\alpha} D_{cph}({\bf q},i\omega_{\nu})\nonumber\\
  &+& 4\alpha D_{c}({\bf q},i\omega_{\nu}).
\end{eqnarray}
Here the intrinsic phonon part $D_{ph}({\bf q},i\omega_{\nu})$
is the Fourier transform of
$-\langle T_{\tau} \phi_{\bf q}(\tau)\phi_{-{\bf q}}
 \rangle_{\bar{H}}$ with the subscript $\bar{H}$ representing
that the $\tau$ dependence of $\phi_{\bf q}$ as well as the
thermodynamic average should be considered with respect to $\bar{H}$.
The charge part $D_{c}({\bf q},i\omega_{\nu})$ and the charge-phonon
coupling part $D_{cph}({\bf q},i\omega_{\nu})$ are, respectively,
obtained by the Fourier transform of
$-\langle T_{\tau} \rho_{\bf q}(\tau)\rho_{-{\bf q}}
\rangle_{\bar{H}}$ and $-\langle T_{\tau}
(\phi_{\bf q}(\tau)\rho_{-{\bf q}}
+\rho_{\bf q}(\tau)\phi_{-{\bf q}})\rangle_{\bar{H}}$,
where the density operator $\rho_{\bf q}$ is defined as
$\rho_{\bf q} = N^{-1/2}
\sum_i e^{i{\bf q} \cdot {\bf R}_i} \rho_i$. 

In terms of a complete set of energy eigenstates $\{|m\rangle\}$
satisfying $\bar{H} |m\rangle = E_m |m\rangle$, we can give the
spectral representation for these quantities.
For example, $D_{ph}({\bf q},i\omega_{\nu})$ is expressed as
\begin{eqnarray}
  \label{eq:phonongf}
  D_{ph}({\bf q},i\omega_{\nu}) &=& \int_0^{\infty} {2 \over \pi}
  d\Omega {\Omega \over (i\omega_{\nu})^2 - \Omega^2}
  B_{ph}({\bf q},\Omega),
\end{eqnarray}
where the phonon spectral function $B_{ph}$ is defined as
\begin{eqnarray}
  \label{eq:phononsp}
  B_{ph}({\bf q},\Omega) &=& \pi (1-e^{-\Omega/T})
  {1 \over \sum_m e^{-E_m/T}} \sum_{nm} e^{-E_m/T} \nonumber\\
  &\times & |\langle m|\phi_{\bf q}|n\rangle |^2
  \delta(\Omega-E_n+E_m).
\end{eqnarray}
At $T=0$, we need to consider only the ground state $|0\rangle$
in the $m$ sum in Eq.~(\ref{eq:phononsp}) and $\tilde{\omega}_0$
can be calculated by the difference between $E_0$ and $E_n$
for excite states, $|n\rangle$'s,
satisfying $\langle 0|\phi_{\bf q}|n\rangle \neq 0$.
Thus our problem is reduced to the analysis of energy eigenstates.

\subsection{Energy spectrum structure and second unitary
  transformation}
\label{sec:3B}

For a further analysis of the HH model in the antiadiabatic and
strong-coupling region, we need to know the characteristic
energy structure of the model.
The structure can be best explained by starting with the atomic
limit, i.e., $t=0$ so that $\bar{H}=\bar{V}$, and the case of $U>0$.
Irrespective of the electron filling $n$, the total energy of
an energy eigenstate can be specified by a couple of nonnegative
integers $N_{ph}$ and $N_{d}$ which represent, respectively,
the total numbers of phonons and doubly occupied sites in the state.
Of course, a high degree of degeneracy exists for the energy
$\omega_0 N_{ph} + U N_{d} -(\mu+U_{ph})nN$ due to the many
possibilities of choosing sites to place up- or down-spin electrons
and those to excite phonons.
However, the ground-state manifold characterized by
$(N_{ph},N_{d})=(0,0)$ is well separated from the
first-excited-state manifold by the energy of either $\omega_0$
or $U$, whichever smaller.
With the increase of $t$, a ``band'' will be formed to lift the
degeneracy to some extent in each manifold, but as long as $t$ is
small compared to $U$ or $\omega_0$, the pair
$(N_{ph},N_{d})=(0,0)$ remains to be a set of good quantum numbers
for the ground-state ``band''.
The same is true for other excited-state manifolds, if there is no
accidental degeneracy among different sets of $(N_{ph},N_{d})$'s.
(The degeneracy may occur if the ratio $U/\omega_0$ is some
rational number.)
Thus, in general, the present system has a structure for the energy
spectrum such as shown schematically in Fig.~\ref{fig:2}(a).

A similar consideration can be made for the case of $U<0$.
Here the ground-state manifold has the largest number of $N_{d}$
under the given total electron number $nN$.
Then, if we reinterpret $N_{d}$ in the above argument as the reduction
in the number of doubly-occupied sites from the ground-state manifold
and take $|U|$ instead of $U$, we obtain exactly the same energy
spectrum structure as shown in Fig.~\ref{fig:2}(a).

A special treatment is necessary for the case of $U=0$,
or $|U| \lesssim t$ if we take the ``bandwidth'' into account.
This is one of the cases in which the ratio $U/\omega_0$ is
equal to some rational number.
Actually, this has a simpler energy spectrum structure
than before as illustrated in Fig.~\ref{fig:2}(b),
because $N_{d}$ is no longer meaningful and $N_{ph}$
is the only good quantum number here.

The energy eigenstates in each manifold can be described by an
effective Hamiltonian $H_{\rm eff}$ which can be derived from
$\bar{H}$ by eliminating the polaron hopping effects on
inter-manifold processes, i.e., the virtual transitions accompanied
with the change in the energy $\omega_0N_{ph}+|U|N_{d}$.
For the ground-state manifold, a considerable number of literature
have already been accumulated as to the derivation of $H_{\rm eff}$
for the Hubbard model, \cite{anderson1,takahashi,macdonald}
the Holstein model, \cite{freericks2,hirsch}
and the HH model. \cite{beni}
Roughly speaking, all these theories make a degenerate perturbation
expansion with respect to $t$ up to second or fourth order.

We extend such an approach even to the case of excited-state
manifolds.
For the purpose, we transform $\bar{H}$ into $H_{\rm eff}$
by introducing some adequate unitary operator $e^{R}$.
We define the antihermitian operator $R$ perturbatively on the
principle that we retain only such terms as to make the energy
$\omega_0 N_{ph}+|U|N_{d}$ invariant among the polaron hopping
processes in each order of $t$.

Following the principle faithfully, we obtain an expansion form of
$H_{\rm eff}$ up to second order in $t$ as
\begin{equation}
  H_{\rm eff}= \bar{V} + H_{\rm eff}^{(1)}+H_{\rm eff}^{(2)},
\end{equation}
where $H_{\rm eff}^{(1)}$ describes the first-order polaron hopping
processes and $H_{\rm eff}^{(2)}$ represents the polaron-polaron
interaction part induced by second-order virtual hopping processes.
In Fig.~\ref{fig:3}, all the processes contributing to
$H_{\rm eff}^{(2)}$ are shown schematically.
Mathematical details to obtain $H_{\rm eff}^{(1)}$ and
$H_{\rm eff}^{(2)}$ are given in Appendix A.

\subsection{Spin versus polaron liquid}
\label{sec:3D}

According to the Holstein's argument on small polarons,
\cite{holstein}
we readily find that the characteristic energy for 
$H_{\rm eff}^{(1)}$, $\epsilon_1$, is 
the polaron hopping energy $te^{-\alpha}$.
Some analysis is needed to find the energy scale to characterize
$H_{\rm eff}^{(2)}$, $\epsilon_2$.
The matrix elements related to the hopping processes
among adjacent three sites [Fig.~\ref{fig:3}(c)-(e)]
are exponentially small in the antiadiabatic and
strong-coupling region. \cite{beni}
In addition, the matrix element for the process (b),
the bipolaron hopping process between nearest neighbor sites,
is also found to vanish exponentially.
Thus only the process (a) survives.
The estimation of the corresponding matrix element suggests
$\epsilon_2 = t^2/U_{ph}$. \cite{comment}

Comparison of $\epsilon_1$ with $\epsilon_2$ implies two situations:
One is the case of $\epsilon_1 \ll \epsilon_2$, or equivalently
$(U_{ph}/\omega_0)e^{-(U_{ph}/\omega_0)} \ll (t/\omega_0)$,
in which we can neglect $H_{\rm eff}^{(1)}$ and need to treat
only $H_{\rm eff}^{(2)}$, leading to some 
spin-model Hamiltonian for $H_{\rm eff}$.
Thus we call this case ``the spin region''.
The other is the case of $\epsilon_1 \gtrsim \epsilon_2$ in which
$H_{\rm eff}^{(1)}$ contributes much 
to the properties of the system.
We call the latter situation ``the polaron-liquid region''.

Although the condition to reach the polaron-liquid region is
compatible with the condition of antiadiabaticity and
strong coupling [$(t/\omega_0) \ll 1 \ll (U_{ph}/\omega_0)$],
the spin region is realized eventually for extremely large
values of $\alpha = U_{ph}/\omega_0$.
Thus we mainly consider the spin region in relation to the
phonon properties in this paper.
However, the polaron-liquid region, especially its ground state,
is quite intriguing in connection with superconductivity.
Therefore we also give a concrete form of $H_{\rm eff}$
for the polaron-liquid ground-state manifold in Appendix B 
in order to provide a theoretical basis for a future study of
superconductivity.

\subsection{Spin Hamiltonian}
\label{sec:3E}

We now provide a convenient form of $H_{\rm eff}$ to determine
$\tilde{\omega}_0$ for the intrinsic phonon mode in the spin regime.
Note that $\tilde{\omega}_0$ is given through the evaluation of
$B_{ph}({\bf q},\Omega)$ in Eq.~(\ref{eq:phononsp}).
The evaluation can be done using $H_{\rm eff}$ for the manifolds
labeled by $N_{ph}= 0, 1, 2, \cdots$ and $N_{d}=0$.
Thus, although no restriction is imposed on $N_{ph}$, we confine
ourselves to the manifolds specified by $N_{d}=0$.
As for $n$, we consider only the half-filling case for simplicity.
Since a concrete form for $H_{{\rm eff:}N_{ph}}$ the effective
Hamiltonian for the manifold labeled by ($N_{ph},N_{d}=0$)
depends critically on the nature of the ground-state manifold,
we need to treat the three cases, i.e., $U>0$, $U<0$, and $U=0$,
separately.
In the following, we give our final result for
$H_{{\rm eff:}N_{ph}}$ in each case.
Some tips for deriving it are explained in Appendix A.

Let us start with the case of $U>0$ (the SDW region), in which 
each site is occupied by one electron in the manifolds with
$N_{d}=0$.
Suppressing the chemical-potential term, we obtain
\begin{eqnarray}
  \label{eq:spin}
  && H_{{\rm eff:}N_{ph}} = N_{ph} \omega_0 +
  \sum_{\langle i,j \rangle} h(i,j:{\rm SDW}) \nonumber \\ 
  &&\times 2t^2e^{-2\alpha} \sum_{p,q}\sum_{L}
  \frac{(-2\alpha)^p}{p!(p+L)!} 
  \frac{(-2\alpha)^q}{q!(q+L)!} \nonumber \\ 
  &&\times \frac{(2\alpha)^L}{|U|+L \omega_0}
  (a^{\dagger}_{ij})^{p}(a_{ij})^{p+L}
  (a^{\dagger}_{ij})^{q+L}(a_{ij})^q, 
\end{eqnarray}
where the symbol $\sum_{p,q}$ indicates the sum over nonnegative 
integers $p$ and $q$ under the condition of $0 \le p+q \le N_{ph}$
for a given number of $N_{ph}$ and another symbol $\sum_{L}$
specifies the sum over integer $L$ from $-{\rm min}(p,q)$ to infinity
provided that the term $L=-|U|/\omega_0$ should be skipped,
if $|U|/\omega_0$ is an integer.
The spin operator $h(i,j:{\rm SDW})$ is defined as
\begin{equation}
  \label{eq:effSDW}
  h(i,j:{\rm SDW})=
  2 \left ( {\bf S}_i\cdot{\bf S}_j-\rho_i \rho_j/4 \right ),
\end{equation}
with 
${\bf S}_i \cdot{\bf S}_j = 
S_i^z S_j^z +(S_i^+  S_j^- + S_i^- S_j^+)/2$,
where $S_i^z = (n_{i\uparrow}-n_{i\downarrow})/2$, 
$S_i^+=c^{\dagger}_{i\uparrow}c_{i\downarrow}$, and 
$S_i^-=c^{\dagger}_{i\downarrow}c_{i\uparrow}$.
Although the presence of phonons brings about some modification,
it seems to be quite natural to obtain the Heisenberg-like
spin Hamiltonian in this SDW region in which the electron spin
resides at each site.

Difference appears in the case of $U<0$ (the CDW region), in which
the ground-state manifold has the largest number of doubly-occupied.
In this case, we obtain $H_{{\rm eff:}N_{ph}}$ in the form of 
Eq.~(\ref{eq:spin}) with the replacement of the spin operator
$h(i,j:{\rm SDW})$ by $h(i,j:{\rm CDW})$, defined by
\begin{equation}
  \label{eq:effCDW}
  h(i,j:{\rm CDW})= 2 \left (\eta_i^z \eta_j^z-1/4 \right ),
\end{equation}
where $\eta_i^z = (\rho_{i}-1)/2$ is the $z$-component
of pseudospin operators. \cite{yang}
Note that the pseudospin transverse components corresponding to
the bipolaron hopping process [the process (b) in Fig.~\ref{fig:3}]
vanish because of the immobile nature of bipolarons.

At $U=0$, $H_{{\rm eff:}N_{ph}}$ is also obtained in the form 
of Eq.~(\ref{eq:spin}), but $h(i,j:{\rm SDW})$ should be replaced with
$h(i,j:U=0)$, defined by
\begin{eqnarray}
  \label{eq:effU0}
  h(i,j:U=0) &=& h(i,j:{\rm SDW})+2h(i,j:{\rm CDW}) \nonumber\\ 
  &+& (\rho_i+\rho_j)/2 \nonumber \\
  &=& 2({\bf S}_i \cdot {\bf S}_j+\rho_i\rho_j/4)-(\rho_i+\rho_j)/2.
\end{eqnarray}
This expression visualizes perfect symmetry of the system with
$U=0$ between spin and charge fluctuations.
This result at $N_{ph}=0$ was derived in Ref.~\onlinecite{beni}.

With these $U$-dependent spin operators $h(i,j:U)$,
$H_{{\rm eff:}0}$ the Hamiltonian to describe the ground-state
manifold is reduced further as
\begin{eqnarray}
  \label{eq:eff:0}
  H_{{\rm eff:}0}= J_0  \sum_{\langle i,j \rangle}h(i,j:U),
\end{eqnarray}
where $J_0$ is defined as
\begin{equation}
  \label{eq:j0}
  J_0(U_{ee},U_{ph}) = 2t^2e^{-2\alpha} \sum_{\ell=0}^{\infty}
  \frac{(2\alpha)^{\ell}}{\ell !} \frac{1}{|U|+\ell \omega_0},
\end{equation}
for $U \neq 0$.
At $U=0$, the $\ell$ sum starts with $\ell=1$.
For large $\alpha = U_{ph}/\omega_0$, we can evaluate the $\ell$
sum easily and obtain
\begin{equation}
  \label{eq:j00}
  J_0(U_{ee},U_{ph}) = \frac{2t^2}{|U_{ee}-2U_{ph}|+2U_{ph}}.
\end{equation}
For $U>0$ or $U_{ee}>2U_{ph}$, this is reduced to $2t^2/U_{ee}$,
the Anderson's super-exchange interaction in the Hubbard model.
\cite{anderson1}

\section{Phonon energy shift}
\label{sec:4}

\subsection{Ground state}
\label{sec:4A}

With reference to the conventional theories on the spin Hamiltonians,
we readily know the ground state $|0\rangle$ of $H_{{\rm eff:}0}$.
For $U>0$, the antiferromagnetic (AF) state in the Heisenberg model
appears as the ground state with $E_0$ the ground-state energy
depending on the lattice structure.
In general, $E_0$ per site is given as
\begin{equation}
  \label{eq:E0}
  E_0/N = -zJ_0 S \left ( S +\delta \right) -zJ_0/4,
\end{equation}
where $S(=1/2)$ is the spin magnitude and $\delta$ represents the
amount of spin fluctuations in the AF state.
In the two-site case with $z=1$, $\delta$ is equal to unity.
Another exact result is known for the one-dimensional (1D) chain with
$z=2$, in which $\delta$ is equal to $2{\rm ln}2-1=0.386$.\cite{bethe} 
For higher dimensions, $E_0$ is not known exactly, but the spin-wave
theory provides reliable estimates for $\delta$.\cite{anderson2,kubo}
For 2D square [$z=4$], simple cubic [$z=6$], and body-centered cubic
(bcc) [$z=8$] lattices, 
the values of $\delta$ have been estimated as $0.158$, $0.097$, and
$0.073$, respectively.

For $U<0$, the ``AF'' state also appears as the ground state,
but ``spin'' fluctuations are absent due to the Ising nature.
Thus we obtain $E_0$ in the form of Eq.~(\ref{eq:E0})
with $\delta=0$ irrespective of the lattice structure.
Physically, the present ``AF'' state represents the bipolaron-ordered
bipartite lattice (the CDW state) and the absence of
``spin'' fluctuations reflects immobility of the bipolarons.

At $U=0$, we should consider the competition of these two states.
Although the energy difference is small, it turns out that
the Heisenberg-type AF state appears as the ground state
with the same amount of spin fluctuations as that for $U>0$.
This is because the transverse spin components in the effective
Hamiltonian for $U=0$ are just equal to those for $U>0$.
Thus $E_0$ is given in the form of Eq.~(\ref{eq:E0}) with
$J_0$ at $U_{ee}=2U_{ph}$ in Eq.~(\ref{eq:j0}) and the same $\delta$
depending on the lattice structure.

\subsection{One-phonon excited states: Numerical study}
\label{sec:4B}

The effective Hamiltonian for the manifold specified by
$N_{ph}=1$ and $N_{d}=0$ is given by
\begin{eqnarray}
  \label{eq:eff:1}
  H_{{\rm eff:}1} &=& \omega_0 
  + \sum_{\langle i,j \rangle}h(i,j:U) \nonumber\\
  &\times&
  [ J_0 + J_1  (a_i^{\dagger}-a_{j}^{\dagger})(a_i-a_{j})/2 ],
\end{eqnarray}
with $J_1$, defined as
\begin{eqnarray}
\label{eq:j1}
 J_1(U_{ee},U_{ph}) &=& 2t^2e^{-2\alpha}
  \biggl \{ {2\alpha \over |U|-\omega_0}
   +\sum_{\ell=0}^{\infty} \frac{(2\alpha)^{\ell}}{\ell !}
\nonumber\\
&&\times
\left [ {(2\alpha-\ell-1)^2 \over \ell+1} -1\right ]
\frac{1}{|U|+\ell \omega_0} \biggr \},
\end{eqnarray}
for $U \neq 0$.
At $U=0$, $J_1$ should be evaluated with the $\ell$ sum
starting with $\ell=1$ instead of $\ell=0$.
In the strong-coupling limit, $J_1$ is reduced to
\begin{equation}
\label{eq:j11}
  J_1(U_{ee},U_{ph}) = \frac{t^2}{\alpha U_{ph}}
  \Biggl(\frac{2U_{ph}}{|U_{ee}-2U_{ph}|+2U_{ph}}\Biggr)^3.
\end{equation}
Comparison between Eqs.~(\ref{eq:j00}) and (\ref{eq:j11})
indicates that $J_1/J_0$ is small, i.e.,
of the order of $1/\alpha$.

Now we investigate the intrinsic phonon mode which appears
as a peak in the spectral function $B_{ph}({\bf q},\Omega)$
in Eq.~(\ref{eq:phononsp}).
The one-phonon excited state $|n \rangle$ giving rise to the peak
should not only be an eigenstate of $H_{{\rm eff:}1}$ but also
satisfy the property of
\begin{equation}
  \label{eq:condition}
  \langle 0 | e^{R}\phi_{\bf q}e^{-R} | n\rangle \approx 1.
\end{equation}
In pursuit of such an eigenstate,
we have examined all the eigenstates of $H_{{\rm eff:}1}$
for small-size 1D chains with the periodic boundary condition
by the exact diagonalization method.
This examination clarifies that there exists one eigenstate
$|n(q)\rangle$ satisfying Eq.~(\ref{eq:condition})
for each $q$.
(Other eigenstates give spectral amplitudes of the order of
at most $t^2$.)
Using the corresponding energy eigenvalue $E_n(q)$,
the energy shift for the intrinsic phonon is given by
$\delta \omega_0(q) = \omega_0+E_0-E_n(q)$.

In Fig.~\ref{fig:4}, an example of the calculated results
for $\delta \omega_0(q)$ in units of $J_0$ is given for
the 1D chain up to ten sites.
We observe some characteristic features for 
$\delta \omega_0(q)$ from this calculation:
(1)For $U<0$, the size dependence for $\delta \omega_0(q)$
is negligible, suggesting that this quantity is determined only by
the processes between nearest-neighbor sites.
(2)For $U \geq 0$, the size dependence is appreciable, indicating
the importance of the longer-range fluctuation effects.
(3)The SDW regime ($U\geq 0$) has a larger $\delta \omega_0(q)$
than the CDW regime ($U<0$) for any value of $q$, implying that
the former provides larger density fluctuations than the latter.
A physical interpretation of this result is the following:
The amount of spin fluctuations is larger than that of
charge fluctuations in the strong-coupling region.
Now, for the high-energy phonons, even spin fluctuations can be
detected as a kind of density fluctuations of electrons.
Thus the SDW phase gives larger $\delta \omega_0(q)$'s
than the CDW phase.
(4)Irrespective of the sign of $U$, the function form of
$\delta \omega_0(q)$ is seen to be $1-\cos q$.

\subsection{Asymptotically exact expression}
\label{sec:4C}

A closer examination of the above numerical study of
$H_{{\rm eff:}1}$ reveals that we can write $|n({\bf q})\rangle$ as
\begin{equation}
  \label{eq:awfn}
  |n({\bf q}) \rangle = a_{\bf q}^{\dagger}|0 \rangle, 
\end{equation}
with $|0\rangle$ the ground state of $H_{{\rm eff:}0}$.
Although this is an approximation to the true eigenstate of
$H_{{\rm eff:}1}$ for a finite value of $\alpha$, this expression
becomes exact in the strong-coupling limit, because 
$J_1$ becomes very small compared to $J_0$.

With use of Eq.~(\ref{eq:awfn}), $E_n({\bf q})$ is obtained as
\begin{eqnarray}
  \label{eq:effq}
  E_n({\bf q}) &=& \omega_0 + E_0 -\delta \omega_0({\bf q}),
\end{eqnarray}
where $\delta \omega_0({\bf q})$ the phonon energy shift is given by
\begin{eqnarray}
\label{eq:phshift}
  \delta \omega_0({\bf q})= - J_1
   {E_0 \over NJ_0} \frac{1}{z}\sum_{\bf a}
   (1-\cos {\bf q} \cdot {\bf a}),
\end{eqnarray}
with ${\bf a}$ a vector connecting nearest neighbor sites.

Using Eqs.~(\ref{eq:E0}) and (\ref{eq:j11}),
we can rewrite Eq.~(\ref{eq:phshift}) further as
\begin{eqnarray}
  \label{eq:shift}
  {\delta \omega_0({\bf q}) \over \omega_0} &=& 
  {1+\delta \over 2}\left ( {t \over U_{ph}} \right )^2
  \Biggl(\frac{2U_{ph}}{|U_{ee}-2U_{ph}|+2U_{ph}}\Biggr)^3
  \nonumber\\ &&\times \sum_{\bf a}(1-\cos {\bf q} \cdot {\bf a}).
\end{eqnarray}
This is our final result and it becomes exact in the strong-coupling
limit.
As explained in Sec.~\ref{sec:4A}, $\delta$, a key quantity in this
expression, depends on the nature of the ground state as well as the
lattice structure. 
Specifically, it provides the jump in $\delta \omega_0({\bf q})$
at the CDW-SDW phase transition with the increase of $U_{ee}$.
Note that this analytic result agrees quantitatively well with the
numerical estimate for $\delta \omega_0(\pi)$ in a two-site
system in Sec.~\ref{sec:2}.
We also note that $\delta \omega_0({\bf q})$ decreases monotonically
with the increase of $|U_{ee}-2U_{ph}|$.
The quantity $|U_{ee}-2U_{ph}|$ represents the charge excitation gap
in the atomic limit \cite{hotta2}
and its increase suppresses the density fluctuations, leading to
the decrease of $\delta \omega_0({\bf q})$.

\section{Conclusion and Discussion}
\label{sec:5}

In summary, we have obtained an asymptotically exact expression for
the renormalized energy of optic phonons, $\tilde{\omega}_0({\bf q})$, 
in the strong-coupling and antiadiabatic limit of the HH model by
using the unitary transformation techniques to derive $H_{\rm eff}$.
A jump structure in $\tilde{\omega}_0({\bf q})$ is found at the
CDW-SDW regime boundary with a larger phonon energy shift in the SDW
region.

We have considered the insulating phase at $T=0$, but a little
different result will be obtained if $T$ is low but larger than the
bandwidth of the ground-state manifold, i.e., $J_0$.
We treat the case in Appendix C.

Our calculation has been done for the half-filled band.
If the exchange interactions, $J_0$ and $J_1$
in Eqs.~(\ref{eq:eff:0}) and (\ref{eq:eff:1}), are independent
of the electron filling $n$ as is the case for $U=0$,
a similar calculations can be done rather easily for
$\delta \omega_0({\bf q})$ off the half-filling
and we may be able to obtain the $n$ dependence such as
$\delta \omega_0({\bf q}) \propto n(2-n)$ as suggested in AC.
However, in general, these exchange interactions depend on $n$,
or more specifically the structure of the state realized with
these exchange interactions.
This provides a highly complicated self-consistent problem
which we have no idea to solve at present.

\acknowledgments

We are grateful to T. Higuchi for useful discussions at the early
stage of this research. 
We also thank the Supercomputer Center, Institute for Solid State 
Physics, University of Tokyo for the facilities and the use of the 
FACOM VPP500. 
One of the authors (TH) has been supported by the Grand-in-Aid for
Encouragement of Young Scientists from the Ministry of Education,
Science, Sports, and Culture (ESSC) of Japan. 
Another (YT) is partly supported by the Grand-in-Aid for Scientific
Research on Priority Areas, ``Research on Molecular Conductors'' as
well as for Scientific Research (C) under the contract 
No.~08640447 from the Ministry of ESSC of Japan.
YT also acknowledges the support for the study of 
organic conductors from the Mitsubishi Foundation.

\appendix

\section{Derivation of the Effective Hamiltonian}
\label{sec:A}

This Appendix presents mathematical details to obtain
the effective Hamiltonian $H_{\rm eff}$ from $\bar{H}$
in Eq.~(\ref{eq:Hbar}).
We transform $\bar{H}$ unitarily as
$H_{\rm eff}= e^{R} \bar{H} e^{-R}$, where $R$ is the antihermitian
operator, determined perturbatively using the guiding principle that
we select only such terms as to make $\omega_0 N_{ph}+|U|N_{d}$
invariant among the polaron hopping processes in each order in $t$.

Before we consider $e^{R}$ in detail, let us decompose
$\bar{T}$ the hopping part of $\bar{H}$ into two parts as
\begin{equation}
  \bar{T} = T'+T'',
\end{equation}
where $T'$ makes $\omega_0 N_{ph}+|U|N_{d}$ invariant,
while $T''$ variant.
In order to obtain more concrete expressions for $T'$ and $T''$,
consider the polaron hopping part between $i$ and $j$ sites,
$\sum_{\sigma} c_{i\sigma}^{\dagger}c_{j\sigma}$,
in Eq.~(\ref{eq:hopping}) first.
In line with Ref.~\onlinecite{macdonald}, we separate the part
into three components as
\begin{eqnarray}
  \label{eq:hop}
  \sum_{\sigma} c_{i\sigma}^{\dagger}c_{j\sigma} =
  T_0(i,j)+T_1(i,j)+T_{-1}(i,j).
\end{eqnarray}
Here $T_0$ represents the processes to make $N_d$ invariant,
given by
\begin{eqnarray}
  T_0(i,j)=\sum_{\sigma}
  (n_{i-\sigma}c_{i\sigma}^{\dagger}c_{j\sigma}n_{j-\sigma}+
  h_{i-\sigma}c_{i\sigma}^{\dagger}c_{j\sigma}h_{j-\sigma}),
\end{eqnarray}
with $h_{i\sigma} \equiv 1-n_{i\sigma}$.
On the other hand, $T_1(i,j)$ and $T_{-1}(i,j)$ are,
respectively, given by
\begin{equation}
  T_1(i,j) = \sum_{\sigma}
  n_{i-\sigma}c_{i\sigma}^{\dagger}c_{j\sigma}h_{j-\sigma},
\end{equation}
and 
\begin{equation}
  T_{-1}(i,j) = \sum_{\sigma}
  h_{i-\sigma}c_{i\sigma}^{\dagger}c_{j\sigma}n_{j-\sigma}.
\end{equation}
The former (latter) describes the process which increases (decreases)
$N_d$ by unity.

Now we decompose the phonon part, $X_i^{\dagger}X_j$, in a similar
way, but care should be taken as for the condition to make the energy
$\omega_0 N_{ph}+|U|N_{d}$ invariant.
Since the polaron hopping process is associated with the energy change
by $mU$ with $m=0$ or $\pm 1$, we need to separate 
$X_i^{\dagger}X_j$ into the component $\Phi_k(i,j)$ making
the change in phonon number by $k=-mU/\omega_0$ and the remainder
$\bar{\Phi}_k(i,j)$:
\begin{equation}
  \label{eq:phonon}
  X_i^{\dagger}X_j = \Phi_k(i,j)+\bar{\Phi}_k(i,j).
\end{equation}
Instead of $\Phi_k(i,j)$, we give the definition for $\bar{\Phi}_k$
explicitly as
\begin{eqnarray}
  \label{eq:phi}
  \bar{\Phi}_k(i,j) &=& e^{-\alpha} {\sum_{\ell_1,\ell_2}}^{(k)}
  \frac{(\sqrt{2\alpha})^{\ell_1}(-\sqrt{2\alpha})^{\ell_2}}
  {\ell_1!\ell_2!} \nonumber \\ 
  &\times& (a_{ij}^{\dagger})^{\ell_1}(a_{ij})^{\ell_2},
\end{eqnarray}
where $a_{ij} \equiv (a_i-a_j)/\sqrt{2}$ and
$\sum_{\ell_1,\ell_2}^{(k)}$ denotes the sum over
$\ell_1(\ge 0)$ and $\ell_2(\ge 0)$ except for $\ell_1-\ell_2=k$.
Note that for noninteger $k=\pm U/\omega_0$, $\bar{\Phi}_k(i,j)$
is nothing but $X_i^{\dagger}X_j$ itself.

Combining Eqs.~(\ref{eq:hop}) and (\ref{eq:phonon}), we obtain
$T'$ and $T''$ as
\begin{equation}
  T' = \sum_{m=\pm1,0}T'_m,~~~~~ T'' = \sum_{m=\pm1,0}T''_m,
\end{equation}
where $T'_m$ and $T''_m$ are, respectively, given by
\begin{equation}
  \label{eq:tmp}
  T'_m = - t \sum_{i,j} N_{ij} T_m(i,j) \Phi_{-mU/\omega_0}(i,j),
\end{equation}
and
\begin{equation}
  \label{eq:tm}
  T''_m = - t \sum_{i,j} N_{ij} T_m(i,j)
   \bar{\Phi}_{-mU/\omega_0}(i,j). 
\end{equation}
If $U/\omega_0$ is not an integer, both $T'_1$ and $T'_{-1}$ vanish.

We now perform an canonical transformation of $\bar{H}$ with the
unitary operator $e^{R}$ as $H_{\rm eff}=e^{R} \bar{H} e^{-R}$
and we rewrite $H_{\rm eff}$ in powers of $R$ as
\begin{eqnarray}
  \label{eq:unitary}
  H_{\rm eff} = \bar{H} + [R,\bar{H}] 
  + \frac{1}{2!}[R, [R,\bar{H}]] + \cdots, 
\end{eqnarray}
where $[A,B] \equiv AB-BA$. 
We expand $H_{\rm eff}$ and $R$ in powers of $t$ as
\begin{equation}
  H_{\rm eff} = \sum_{j=0}^{\infty} H_{\rm eff}^{(j)}
  ~~{\rm and}~~
  R = \sum_{j=1}^{\infty} R^{(j)}.
\end{equation}
Note that $H_{\rm eff}^{(0)}=\bar{V}$.

In first order, if we determine $R^{(1)}$ to satisfy
\begin{equation}
  \label{eq:s1}
  [\bar{V}, R^{(1)}] = T'',
\end{equation}
$H_{\rm eff}^{(1)}$ is simply given by 
\begin{equation}
  \label{eq:h1}
  H_{\rm eff}^{(1)}= T'.
\end{equation}
The solution to Eq.~(\ref{eq:s1}) is written as
\begin{equation}
  R^{(1)}=R_0^{(1)}+R_1^{(1)}+R_{-1}^{(1)}, 
\end{equation}
with $R_m^{(1)}$, expressed as
\begin{equation}
  \label{eq:sm}
  R_m^{(1)} = - t \sum_{i,j} N_{ij}T_m(i,j) \phi_m(i,j),
\end{equation}
where the operator $\phi_m(i,j)$ is defined through the equation of
\begin{eqnarray}
  mU \phi_m(i,j) &+& \omega_0
  [a_i^{\dagger}a_i+a_j^{\dagger}a_j,\phi_m(i,j)]  \nonumber \\ 
  &=& \bar{\Phi}_{-mU/\omega_0}(i,j). 
\end{eqnarray}
We find an explicit solution to this equation in the following way:
\begin{eqnarray}
  \label{eq:phim}
  \phi_m(i,j) &=& e^{-\alpha}
  {\sum_{\ell_1,\ell_2}}^{(-mU/\omega_0)}
  \frac{(\sqrt{2\alpha})^{\ell_1}(-\sqrt{2\alpha})^{\ell_2}}
  {\ell_1!\ell_2!} \nonumber \\ 
  &\times& \frac{1}{mU+\omega_0(\ell_1-\ell_2)}
  (a_{ij}^{\dagger})^{\ell_1}(a_{ij})^{\ell_2}. 
\end{eqnarray}
The obtained $R^{(1)}$ is antihermitian, i.e.,
$R^{(1)}=-{R^{(1)}}^{\dagger}$.

In second order, using Eq.~(\ref{eq:s1}), we can write
\begin{eqnarray}
  H_{\rm eff}^{(2)}= [R^{(1)},T']
 + [R^{(1)},T'']/2 +[R^{(2)},\bar{V}]. 
\end{eqnarray}
By determining $R^{(2)}$ so as to eliminate the terms
associated with the change in $\omega_0 N_{ph} +|U| N_{d}$,
namely, choosing $R^{(2)}$ satisfying
\begin{eqnarray}
\label{eq:s2}
   [\bar{V}, R^{(2)}] = [R^{(1)},T']+[R^{(1)},T'']''/2,
\end{eqnarray}
we can reduce $H_{\rm eff}^{(2)}$ into the following form:
\begin{eqnarray}
  \label{eq:h2}
  H_{\rm eff}^{(2)} =[R^{(1)},T'']'/2,
\end{eqnarray}
where the symbols prime and double prime on the square brackets
in Eqs.~(\ref{eq:s2}) and (\ref{eq:h2}) indicate that
we should include only the processes making the energy
$\omega_0 N_{ph} + |U| N_{d}$ invariant and variant, respectively.

 From these compact expressions for $H_{\rm eff}^{(1)}$ and 
$H_{\rm eff}^{(2)}$, the forms more convenient
for actual calculations can be derived.
Especially in the spin region, we obtain $H_{\rm eff}$'s as given
in Sec.~\ref{sec:3E}.
The process to reach these $H_{\rm eff}$'s are straightforward,
but a care should be taken in the reduction of
$H_{\rm eff}^{(2)}$:
In general, to the process (a) in Fig.~\ref{fig:3}, three terms out of
the expression $[R^{(1)},T'']'/2$, i.e., $[R_0^{(1)}, T''_0]'$,
$(R_{-1}^{(1)}T''_{1}-T''_{-1}R_{1}^{(1)})'$, and
$(R_{1}^{(1)}T''_{-1}-T''_{1}R_{-1}^{(1)})'$ contribute. 
However, in the case of $U>0$, only the second term
$(R_{-1}^{(1)}T''_{1}-T''_{-1}R_{1}^{(1)})'$ counts due to
the fact that each site is occupied by at most one electron
in the manifolds with $N_{d}=0$ in the SDW regime.
On the other hand, in the case of $U<0$, the ground-state manifold has
the largest number of doubly-occupied 
sites and the condition of ``$N_{d}=0$'' in this case means that this 
continues to be the case for the excited-state manifolds under
consideration.
Thus instead of the second, the third term 
$(R_{1}^{(1)}T''_{-1}-T''_{1}R_{-1}^{(1)})'$ contributes
to the process (a).
At $U=0$, no correlation works between electrons with
different spins and all the three terms,
$[R_0^{(1)}, T''_0]'$,
$(R_{-1}^{(1)}T''_{1}-T''_{-1}R_{1}^{(1)})'$, and
$(R_{1}^{(1)}T''_{-1}-T''_{1}R_{-1}^{(1)})'$, contribute.

\section{Polaron-liquid Hamiltonian}
\label{sec:B}

As mentioned in Sec.~\ref{sec:3D}, we can think of the region
in the intermediate strength of $\alpha$ in which the polaron
kinetic term $H_{\rm eff}^{(1)}$
cannot be neglected in comparison with $H_{\rm eff}^{(2)}$.
In particular, we already know from the exact-diagonalization
study on the small-size HH model \cite{takada1} as well as
the perturbation-theoretic study on the infinite-dimensional
HH model \cite{hotta} that ``the metallic state'' appears
at $|U| \approx 0$, i.e., the transition region from CDW to SDW.
The state is very interesting from various points of view including
superconductivity in the alkali-metal-doped fullerenes.
\cite{takada3}
Thus in this Appendix, we give $H_{\rm eff}$ for the ground-state
manifold pertinent to the investigation of this polaron-liquid state.

At $U=0$, the second-order part of $H_{\rm eff}$ has already been
given in Sec.~\ref{sec:3E}.
For the ground-state manifold with $N_{ph}=0$, the exchange
interaction $J_0$ can be provided in a more compact fashion
than in Eq.~(\ref{eq:spin}) as
\begin{equation}
\label{eq:j}
  J_0 = {2t^2e^{-2\alpha} \over \omega_0} \sum_{\ell=1}^{\infty}
  \frac{(2\alpha)^{\ell}}{\ell \times \ell !}.
\end{equation}
As for the first-order part, we should include all the hopping
terms, $T'_0$, $T'_1$, and $T'_{-1}$, given in Eq.~(\ref{eq:tmp}),
because double occupancy is permitted at every site due to the
vanishment of a charge-excitation energy.
The sum of these three terms recovers the free polaron hopping
term with the transfer integral $te^{-\alpha}$.

When $|U|$ is not exactly zero but small, i.e., at most of the
order of $t$, we should add the on-site interaction term $\bar{V}$
in Eq.~(\ref{eq:on-site}) as it is to $H_{\rm eff}$ as an additional
perturbation, because the correction to the
operator $n_{i\uparrow}n_{i\downarrow}$ due to the unitary
transformation of $e^{R}$ is at most of the order of $t^2$
in the ground-state manifold.
Combining all these terms, we obtain $H_{\rm eff}$ as 
\begin{eqnarray}
  H_{\rm eff} &=& 
  -t e^{-\alpha} \sum_{\langle i,j \rangle,\sigma}
  (c_{i\sigma}^{\dagger}c_{j\sigma}+{\rm h.c.})
  + U \sum_{i}n_{i\uparrow}n_{i\downarrow} \nonumber \\ 
  &+& 2J_0 \sum_{\langle i,j \rangle}
  \bigl( {\bf S}_i \cdot {\bf S}_j + \frac{1}{4}\rho_i\rho_j \bigr). 
\end{eqnarray}
Note that this expression is valid even for $n$ off the half-filling,
because $J_0$ is determined only through the phonon effect which is
independent of $n$.
We also note that the nearest-neighbor Coulomb interaction term with
the strength $V$ at most of the order of $t$, if any in the original
model Hamiltonian, can also be added to this Hamiltonian as it is.

We shortly comment on the interesting case in which
$U/\omega_0$ happens to be equal to some nonzero integer, say, unity.
In this situation, one phonon-excited and charge-excited states are
degenerate. 
Thus, in constructing $H_{\rm eff}$ to describe the first-excited 
manifold, we should take account of all the one-polaron hopping terms,
$T'_m$ with $m=0$ and $\pm1$, corresponding to the polaron hopping
accompanied with creation or annihilation of local phonons. 
Then we can expect a polaron-mobile excited state in the system of an
insulating ground state. 

\section{Finite-temperature Green's-function approach}
\label{sec:C}

In this Appendix, we consider the temperature range
of $J_0 \alt T \ll \omega_0$ and apply the Green's-function method
to the original Hamiltonian, Eq.~(\ref{hh}).
In this temperature range, the electronic long-range order
originating from the exchange interaction $J_0$ is destroyed
by thermal fluctuations.
The incoherent nearest-neighbor electron-hopping processes
can be treated perturbatively with respect to $t$ from the atomic
$t=0$ limit.

 From the diagrammatic consideration, the phonon Green's function in 
Eq.~(\ref{eq:phD}) is rewritten as 
\begin{equation}
  D({\bf q},i\omega_{\nu})
  =\frac{2\omega_0}
  {(i\omega_{\nu})^2-\omega_0^2
    \left[1-\frac{2\alpha\omega_0\Pi({\bf q},i\omega_{\nu})}
      {1+(U_{ee}/2)\Pi({\bf q},i\omega_{\nu})}\right]}, 
\end{equation}
where $\Pi$ is the proper polarization function. 
Since the renormalized phonon energy $\tilde{\omega}_0$ is determined
by the pole of $D^{\rm R}({\bf q},\omega)$, 
$\tilde{\omega}_0$ is given by the self-consistent
solution of
\begin{equation}
 \label{eq:omega}
  \tilde{\omega}_0^2=\omega_0^2
  \Bigl[
  1-\frac{2\alpha\omega_0 \Pi^{\rm R}({\bf q},\tilde{\omega}_0)}
  {1+(U_{ee}/2)\Pi^{\rm R}({\bf q},\tilde{\omega}_0)}
  \Bigr], 
\end{equation}
where the retarded polarization function is denoted
by $\Pi^{\rm R}$.
Thus our task is to evaluate $\Pi^{\rm R}$ in the antiadiabatic 
and strong-coupling ($U_{ph} \gg \omega_0$) limit.

In terms of real-space representation for the polarization
function $\Pi_{ij}(i\omega_{\nu})$, 
$\Pi({\bf q},i\omega_{\nu})$ is expressed as 
\begin{eqnarray}
\label{eq:A1}
  \Pi ({\bf q},i\omega_{\nu}) = - \frac{1}{N}
  \sum_{i \ne j} [1 - e^{-i{\bf q}\cdot ({\bf R}_i - {\bf R}_j)}]
  \Pi_{ij}(i\omega_{\nu}),
\end{eqnarray}
where we have used the relation of
$\Pi({\bf 0},i\omega_{\nu})=0$ for $\omega_{\nu} \ne 0$.
Up to second order in $t$, only the polarization function
between nearest-neighbor sites $\Pi_{\bf a}$ contributes to
Eq.~(\ref{eq:A1}).
Thus we obtain
\begin{eqnarray}
  \label{eq:pol}
  \Pi ({\bf q},i\omega_{\nu}) = - \sum_{\bf a} 
  (1 - \cos {\bf q}\cdot {\bf a}) \Pi_{\bf a}(i\omega_{\nu}), 
\end{eqnarray}
with
\begin{eqnarray}
  \label{eq:pi1}
  \Pi_{\bf a} (i\omega_{\nu}) &=& -2t^2 T\sum_{n}
  [G(i\epsilon_n)G(i\epsilon_n+i\omega_{\nu}) \nonumber \\ 
  &\times& \Lambda(i\epsilon_n,i\epsilon_n+i\omega_{\nu})]^2,
\end{eqnarray}
where the factor $2$ in the righthand side accounts for the
spin sum, $\epsilon_n\equiv (2n+1)\pi T$ is the fermion Matsubara
frequency with an integer $n$, $G$ and $\Lambda$ are,
respectively, the one-electron Green's function and the three-point
scalar vertex function in the atomic limit.

For $\omega_{\nu} \ne 0$, the Ward-Takahashi identity relates
$\Lambda$ with $G$ as \cite{takada2}
\begin{eqnarray}
  \label{eq:ward}
  \Lambda(i\epsilon_n,i\epsilon_n+i\omega_{\nu}) 
  =\frac{G^{-1}(i\epsilon_n+i\omega_{\nu})-G^{-1}(i\epsilon_n)}
  {i\omega_{\nu}}.
\end{eqnarray}
By substituting Eq.~(\ref{eq:ward}) into Eq.~(\ref{eq:pi1}),
we obtain $\Pi_{\bf a}$ as 
\begin{eqnarray}
  \label{eq:pi1a}
  \Pi_{\bf a} (i\omega_{\nu}) = -4t^2 
  \frac{\chi(i\omega_{\nu})-\chi(0)}{(i\omega_{\nu})^2}, 
\end{eqnarray}
with $\chi(i\omega_{\nu})$, defined by
\begin{eqnarray}
  \chi(i\omega_{\nu}) = -T\sum_{n}
  G(i\epsilon_n)G(i\epsilon_n+i\omega_{\nu}). 
\end{eqnarray}
We can give an analytical expression for $\chi(i\omega_{\nu})$
with use of an exact expression for $G(i\epsilon_n)$ derived
through the Lang-Firsov transformation. \cite{takada2}

For the time being, we assume $|U| \alt T$.
Then $\chi(i\omega_{\nu})$ is reduced into the following form
in the strong-coupling limit at half filling:
\begin{eqnarray}
  \label{eq:chia}
  \chi(i\omega_{\nu}) &=& \frac{(1-f)^2}{2}
  \frac{2U_{ph}+|U|}{(2U_{ph}+|U|)^2-(i\omega_{\nu})^2} \nonumber \\
  &+& f(1-f)
  \frac{2U_{ph}}{(2U_{ph})^2-(i\omega_{\nu})^2} \nonumber \\  
  &+& \frac{f^2}{2}
  \frac{2U_{ph}-|U|}{(2U_{ph}-|U|)^2-(i\omega_{\nu})^2}, 
\end{eqnarray}
with $f \equiv 1/(1+e^{|U|/2T})$.
Although it is not simply given by the limit of
$\omega_{\nu} \to 0$ in Eq.~(\ref{eq:chia}) due to the
presence of a singular term in proportion to $1/T$,
$\chi(0)$ can be well approximated by the value of
$\lim_{\omega_{\nu}\to 0} \chi(i\omega_{\nu})$,
because the weight for the singular term vanishes
exponentially for $U_{ph} \gg \omega_0$.

The analytical continuation of $\Pi({\bf q},i\omega_{\nu})$
in Eq.~(\ref{eq:pol}) together with Eqs.~(\ref{eq:pi1a})
and (\ref{eq:chia}) provides $\Pi^{\rm R}({\bf q},\omega)$.
Because we are interested in the region of
$\omega \approx \omega_0 \ll 2U_{ph}\pm |U|$,
$\Pi^{\rm R}({\bf q},\omega)$ is essentially the same as
$\Pi^{\rm R}({\bf q},0)$.
Thus we can replace $\Pi^{\rm R}({\bf q},\omega)$ by
$\Pi^{\rm R}({\bf q},0)$ in Eq.~(\ref{eq:omega}) and obtain
\begin{equation}
  \delta \omega_0({\bf q})/\omega_0 =
  U_{ph}\Pi^{\rm R} ({\bf q},0),
\end{equation}
up to second order in $t$.
We can rewrite this equation as
\begin{eqnarray}
  \label{eq:shiftA}
  {\delta \omega_0({\bf q}) \over \omega_0} &=& 
  {\eta(T) \over 4} \left ( {t \over U_{ph}} \right )^2
  \Biggl(\frac{2U_{ph}}{|U_{ee}-2U_{ph}|+2U_{ph}}\Biggr)^3
  \nonumber\\
   &&\times \sum_{\bf a}(1-\cos {\bf q} \cdot {\bf a}),
\end{eqnarray}
where $\eta(T)$ describes the $T$ dependence, which
is given by
\begin{eqnarray}
  \label{eq:eta}
  \eta(T) = 1 + 3{|U| \over U_{ph}} f, 
\end{eqnarray}
up to first order in $|U|/U_{ph}$.

We have also considered the case of $|U| \gg T$, starting with
the inspection of $\chi(i\omega_{\nu})$ and obtained exactly the
same expression for $\delta \omega_0({\bf q})$ as in
Eq.~(\ref{eq:shiftA}) with $f$ put equal to zero in
Eq.~(\ref{eq:eta}).

At $U=0$ as well as for $|U| \gg T$, $\eta(T)$ is reduced to unity,
independent of $T$.
Then Eq.~(\ref{eq:shiftA}) provides essentially the same result
for $\delta \omega_0$ as that of AC, \cite{alexandrov2,alexandrov3}
but it is different from that in Eq.~(\ref{eq:shift}) by the factor
of $1/2(1+\delta)$.
This illustrates the subtlety as for the order of two limits,
$U_{ph} \to \infty$ and $T \to 0$.
If we take the $T \to 0$ limit first, the density fluctuations
in the ordered ground state determines $\delta \omega_0({\bf q})$
as given in Eq.~(\ref{eq:shift}).
On the other hand, Eq.~(\ref{eq:shiftA}) corresponds to the result
with taking the $U_{ph} \to \infty$ limit first.
In the latter limit, many degenerate states without the long-range
order contribute to the thermal average even at $T \approx 0$.
The absence of coherent density fluctuations in this case accounts
for the reduction factor $1/2(1+\delta)$
in $\delta \omega_0({\bf q})$.


\begin{figure}
\caption{
(a) An example of the results for excitation energies in the
energy range less than $3\omega_0$ as a function of $U_{ee}$
with a fixed value of $U_{ph}$.
We obtain this result by using an exact diagonalization
method to calculate the pole positions in
the retarded phonon Green's function $D^{\rm R}(q,\omega)$
in a two-site Hubbard-Holstein model at $q=\pi$ in the
antiadiabatic ($t=0.2\omega_0$) and strong-coupling
($U_{ph}=2\omega_0$) case.
There are basically two branches for $\tilde{\omega}_0$.
One is the intrinsic phonon mode (the solid curve) and
the other is the mode of electronic origin (the dashed
curve).
(These modes interchange their nature
at $U_{ee} \approx 5\omega_0$ as indicated by the interchange
of the solid curve with the dashed curve.)
For comparison, the spin-excitation energy is also plotted
by the thin dashed curve.
For $U_{ee} \approx 2U_{ph}$, the third mode (the dotted curve),
the combined charge and phonon mode, cannot be neglected.
(b) Corresponding spectral weight at each pole with the
corresponding species of curve.
The coefficient of delta-function contribution in the
spectral function, $-{1 \over \pi}{\rm Im}D^{\rm R}
(q,\omega) \equiv A_q \delta(\omega -\tilde{\omega}_0(q))$,
is plotted at $q=\pi$. From its definition, the quantity $A_{\pi}$ is
dimensionless.
}
\label{fig:1}
\end{figure}

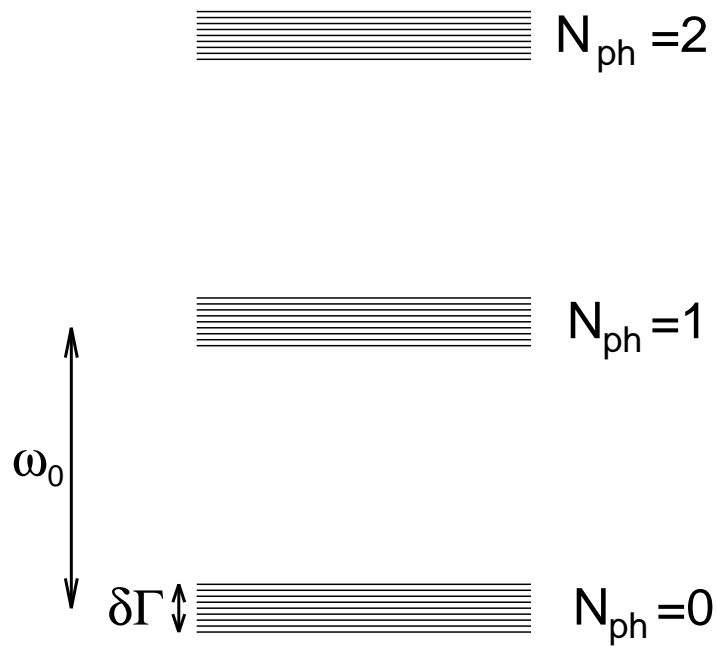
\begin{figure}
\caption{
Schematic representation of the energy spectrum structure
of the Hubbard-Holstein model in the antiadiabatic
($t \ll \omega_0$) and strong-coupling ($U_{ph} \gg \omega_0$)
region.
We can classify two situations.
(a) The CDW ($U<0$) or SDW ($U>0$) case in which $|U|$ is
much larger than $t$.
(b) The charge-spin degenerate case in which $U$ is about
the same as $t$.
}
\label{fig:2}
\end{figure}

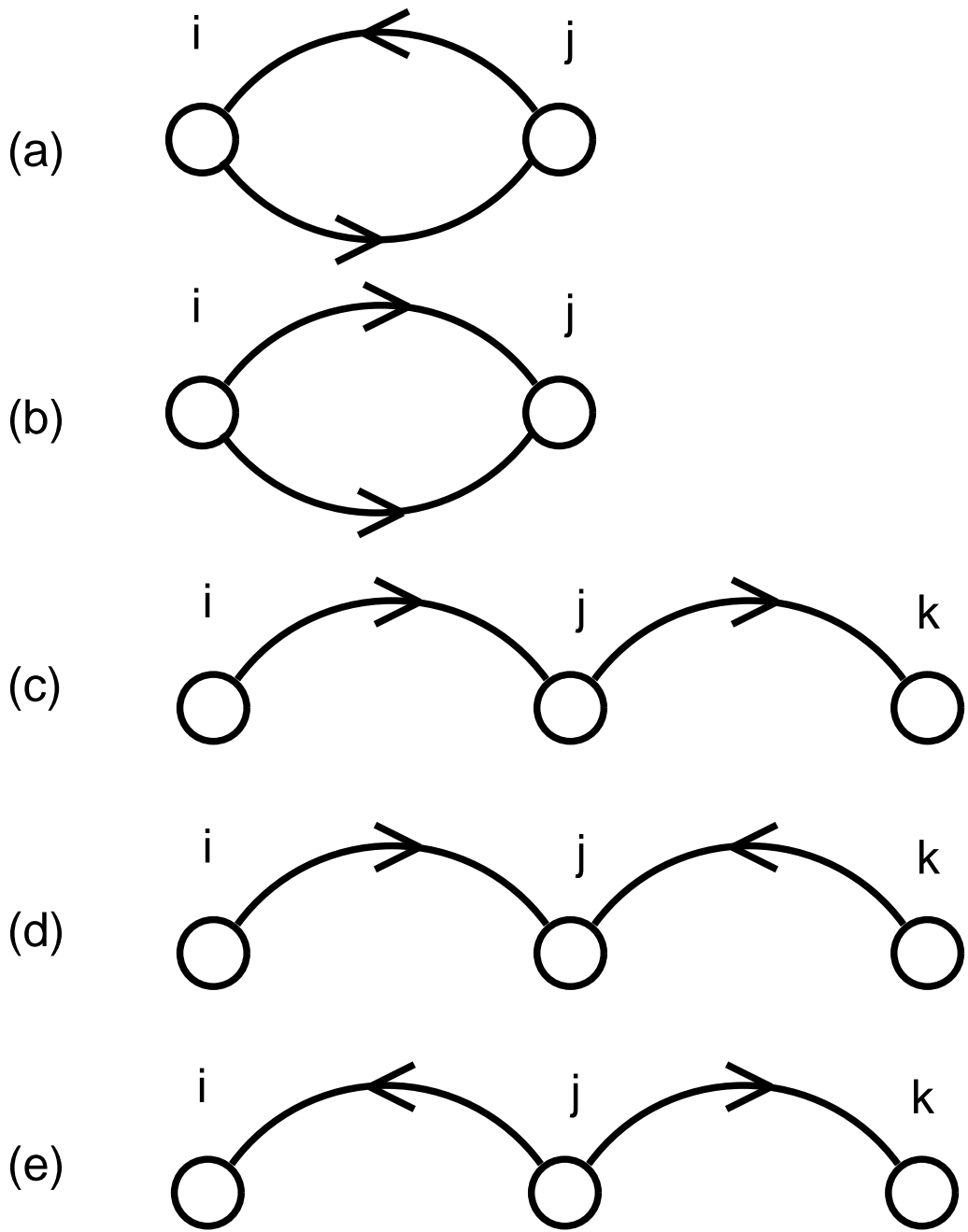
\begin{figure}
\caption{
Second-order polaron hopping processes contributing to
$H_{\rm eff}^{(2)}$.
Two types of the processes between nearest neighbor sites $i$ and $j$
are shown in (a) and (b), while the terms among adjacent three sites
$i$, $j$, and $k$ are represented in (c)-(e). 
}
\label{fig:3}
\end{figure}

\begin{figure}
\caption{
An example of the calculated results for the intrinsic-phonon
energy shift $\delta \omega_0(q)$ in units of $J_0(U)$ in 1D chains
up to ten sites with the periodic boundary conditions.
The open (solid) symbols represent the case of $U<0$ ($U>0$).
The solid and dashed curves show the dependence of $1-\cos q$.
}
\label{fig:4}
\end{figure}

\end{multicols}

\begin{references}
\bibitem{berger2} In the caption of Fig.~8 in 
  Ref.~\onlinecite{berger}, it is stated that $\omega_0=4t$,
  but the behavior of $\tilde{\omega}_0$ in the figure does not
  indicate that this is the case.
  Rather $\omega_0$ should be equal to $t$.
\bibitem{comment} More precisely, $\epsilon_2$ is equal to
  $J_0(U_{ph},U_{ee})$ in Eq.~(\ref{eq:j00}).
  With the change of $U_{ee}$, the mimimum of $J_0(U_{ph},U_{ee})$
  is $t^2/U_{ph}$, which we take as an energy scale for $\epsilon_2$.
\end{references}
\end{document}